\documentclass[twoside,twocolumn]{article}

\usepackage{lmodern}
\usepackage[T1]{fontenc} % Use 8-bit encoding that has 256 glyphs
\usepackage{microtype} % Slightly tweak font spacing for aesthetics
\usepackage[english]{babel} % Language hyphenation and typographical rules
\usepackage{graphicx}%,epstopdf
\graphicspath{{figures/}}
\usepackage{booktabs}

\usepackage{epsfig}

\usepackage[hmarginratio=1:1,top=18mm,bottom=18mm,left=18mm,right=14mm,columnsep=20pt]{geometry}
\usepackage[small,labelfont=bf,up,up]{caption} % hang,textfont=it
\usepackage{booktabs} % Horizontal rules in tables
\usepackage{lettrine} % The lettrine is the first enlarged letter at the beginning of the text

\usepackage{enumitem} % Customized lists
\setlist[itemize]{noitemsep} % Make itemize lists more compact
\usepackage{tabularx,ragged2e}

\usepackage{abstract} % Allows abstract customization

\usepackage{titlesec} % Allows customization of titles
\titleformat{\section}[block]{\large\scshape\centering}{\thesection.}{1em}{} 
\titleformat{\subsection}[block]{\large\scshape\centering}{\thesubsection.}{1em}{}

\usepackage{fancyhdr} % Headers and footers
\usepackage{titling} % Customizing the title section
\usepackage{hyperref} % For hyperlinks in the PDF

\title{\bf Spatial Distribution of Gamma-Ray Burst Sources}
\author{\textsc{\bfseries S. I. Shirokov,$^{1}$}\thanks{E-mail: arhath.sis@yandex.ru} \textsc{\bfseries A. A. Raikov,$^{2}$} and \textsc{\bfseries Yu. V. Baryshev$^{1}$}\\
	\normalsize \itshape $^{1}$St. Petersburg State University, Universitetskii pr. 28, St. Petersburg 198504, Russia\\
	\normalsize \itshape $^{2}$Main (Pulkovo) Astronomical Observatory, Russian Academy of Sciences,\\
	\normalsize \itshape Pulkovskoe shosse d. 65 cor. 1, St. Petersburg 196140, Russia
	\\
}
\date{}
\renewcommand{\maketitlehookd}{%
	\begin{abstract}
		\noindent The spatial distribution of sources of gamma-ray bursts (GRB) with known red shifts is analyzed by the conditional density and pairwise distance methods.  The sample of GRB is based on data from the Swift mission and contains fluxes, coordinates, and red shifts for 384 GRB sources. Selection effects that distort the true source distribution are taken into account by comparing the observed distribution with fractal and uniform model catalogs. The Malmqvist effect is modeled using an approximation for the visible luminosity function of the GRB.  The case of absorption in the galactic plane is also examined.This approach makes it possible to study the spatial structure of the entire sample at one time without artificial truncations. The estimated fractal dimensionality is $D=2.55\pm0.06$ on scales of $2\div6$ Gpc.
		
		{\bf Key words:} gamma-ray bursts: large-scale structure of the universe: fractal dimensionality
	\end{abstract}
}

\begin{document}
	
	\maketitle
	
	\section{Introduction} According to current observational data, the sources of  gamma-ray bursts (GRB) are explosions of massive supernova stars (long GRB) and mergers of neutron stars (short GRB) in distant galaxies. Thus, the spatial distribution of GRB reflects the large-scale distribution of galaxies, so that analyzing the distribution of gamma-ray bursts in space and time is an important task for the study of the evolution of the large-scale structure of the universe. The extreme luminosity of GRB makes it possible to detect their sources at large red shifts, and the available completeness of the surveys of GRB (e.g., Swift) makes it possible to use samples of GRB with known red shifts for preliminary analysis of  their spatial distribution  over a wide range of scales.
	
	The standard cosmological model assumes a uniform distribution of matter in the universe, including dark matter and dark energy. Observations of the spatial distribution of visible matter (galaxies) do, however, reveal nonuniformities over scales much longer than the standard correlation length, 10 Mpc (the Sloan Great Wall, size $\sim300$ Mpc at $z\sim0.07$ [1,2]). In addition, the power law dependence of the conditional density of the distribution of galaxies $\Gamma(r) \propto r^{-1}$ on scales up to 100 Mpc corresponds to a fractal dimensionality $D \sim 2$ [3-5]. Large nonuniformities have recently been discovered in the distribution of galaxies in the SDSS/CMASS survey (the BOSS Great Wall, size $\sim400$ Mpc at $z\sim0.47$ [6]), as well as in the deep galactic survey COSMOS (Super Large Clusters with sizes $\sim 1000$ Mpc at $z\sim1$ [7-9]).
	
	The spatial distribution of gamma-ray bursts has been analyzed in a number of papers [10-13]. Thus, the distribution of 244 GRB has been analyzed [10] as part of the Swift mission using the $\xi$-function method. The correlation length was $r_0\approx388$ h$^{-1}$ Mpc, $\gamma=1.57\pm0.65$(at the $1\sigma$ level), and the uniformity scale was $r\geq7700$ h$^{-1}$. Other approaches to the study of the correlation properties of the spatial structures are the conditional density method [3,4] and the pairwise distance method [14]. In  Ref. 11 it was applied for the first time to 201 GRB with known (at that  time) angular coordinates and red shift. An estimate of $D\cong2.2\div2.5$ was obtained for the fractal dimensionality. This method also makes it possible  to detect close pairs and triplets of points. Thus, for example, a spatially isolated group of five GRB was detected with coordinates $23^h$ $50^m < \alpha < 0^h$ $50^m$ and $5^\circ < \beta < 25^\circ$ a redshift of $0.81 < z < 0.97$. If GRB events are regarded as indicators of the presence of matter in space, then this group is indirect evidence of a supercluster of galaxies within this range  of coordinates.
	
	A giant ring of GRB with a diameter of 1720 Mpc at red shifts of $0.78<z<0.86$ has been discovered [12].The probability that this structure was found randomly is $2 \times 10^{-6}$. 352 GRB have been found [13] to have an estimated dimensionality in terms of the $\Lambda CDM$ model of $D\approx2.3\pm0.1$. In other models, $D\approx2.5$. The latest numerical predictions in terms of the $\Lambda CDM$ model show [15] that at large red shifts it is already possible to observe clusterization of matter, so the detection of structures in the distribution of GRB is an important problem.
	
	In this paper, we estimate fractal dimensionality using the conditional density method for GRB for the first time and compare this estimate with estimates derived from the pairwise distance method. As a comparison with the GRB  catalog, artificial fractal and uniform catalogs are modeled. Effects owing to the sample geometry are taken into account, in particular: limits on the maximum radius sphere and cutoff of the galactic belt. The evolution of luminosity with increasing red shift is examined. For the first time these plots are compared with a uniform distribution, so it is possible to compare the efficiency of the two methods. The power law dependence over a large range of scales also shows up more clearly.
    
    \section{The sample} The Swift Gamma-ray Burst Mission on line catalog [16], which has been extended in Ref. 17 and in another on  line  catalog  [18],  is  used  as  the  basis  for  the  GRB  catalog. Our  catalog  of  GRB  sources  with  known  red  shifts contains  a  total  of  384  objects.    Our  approach  makes  it  possible  to  use  all  the  points  to  determine  the  fractal dimensionality  without  additional  truncations  and  selection. The  only  condition  for  including  objects  in  the combined  catalog  is  the  existence  of  angular  coordinates  and  red  shifts.    Thus,  of  the  working  sample  ($<8$  Gpc), 377 GRB remain, for 360 of which the luminosities have been determined.  The combined catalog has been updated to  June  2017.
    
    The first rows  of our  working  sample are shown  in Figs. 1  and 2.  The headings of the tables correspond to:the  name  of  the  event,  galactic  coordinates,  the  estimated  time  of  the  event,  the  received  radiation  in $10^{-7}$  erg/cm$^{-2}$ in the $15\div150$ keV band during the time $T_{90}$ of the event, the red shift, the mission (Swift), metric coordinates, distance to  the  GRB  source,  the  logarithm  of the  flux,  and  the  logarithm  of  the  luminosity.
    
    \section{Methods} All  the  model  calculations  were  done  using  the  original  Fractal  Dimension  Estimator  program,  which  is described  elsewhere.
    
    \begin{table}[h] \centering \small
	    \caption{Original  Rows  of  the  Catalog.}
		\begin{tabularx}{.48\textwidth}{>{\hsize=0.08\textwidth}rrrrrc} 
		    \hline \hline
			name	&	$l$	&	$b$	&	$T_{90}$	&	$F_{obs}$	&	$z$ \\
			\hline 
			151215A	&	177.25358	&	8.55309		&	17.80	&	3.10	&	2.590	\tabularnewline
			150423A	&	9.70821		&	59.24722	&	0.22	&	0.63	&	1.394	\tabularnewline
			141121A	&	200.39117	&	26.85321	&	549.90	&	53.00	&	1.470	\tabularnewline
			...	\tabularnewline
			\hline			
		\end{tabularx}
		\label{tab:OR}	
    \end{table}

    \begin{table}[h] \centering
		\caption{ Calculated  Rows  of  the  Catalog.}			
		\begin{tabularx}{.48\textwidth}{>{\hsize=0.08\textwidth}rrrrrc} 
			\hline \hline
			$X_{Mpc}$	&	$Y_{Mpc}$	&	$Z_{Mpc}$	&	$R_{Mpc}$	&	$lg S_{obs}$	&	$lg L_b$ \\
			\hline 
			--5862.4	&	281.2   &	882.7	&	5935.1	&	--0.75 	&	51.97	\\
			2099.7 	    &	359.2  	&	3580.2	&	4166.0	&	0.45   	&	52.53	\\
			--3605.5	&	--1340.3&	1947.5	&	4311.5	&	--1.01 	&	51.11	\\
			...	\\
			\hline 			
		\end{tabularx}
		\label{tab:CR}	
    \end{table}

    \subsection{Conditional  density}  This  method,  which  is  discussed  in  detail  in Refs.  3  and  4,  essentially  involves counting the number of points in spheres of different radii.  The conditionality is that the center of a sphere is a point in the set.   For  correct  operation  of  this  method  at large scales, it is  necessary to take the  effect  of  the  boundary  of the  set  into  account  [19].    The  concentration  inside  a  sphere  of  radius  $r$  is  given  by
    
    \begin{equation}
	    n(r) = \frac {1}{N_c(r)}\sum_{i=1}^{N_c(r)} \frac {N_i(r)}{V(r)}, 
    \end{equation}
    
    where $N_c(r)$ is the number of spheres inside the sample, $N_i(r)$ is the number of points in a sphere, and $V(r)$ is the volume of  a  sphere.  On  intermediate  scales  this  distribution  has  a  $D-3$  power  law  dependence,
    
    \begin{equation}
	    \Gamma^*(r)=\langle n(r'<r) \rangle_p 
    \end{equation}	

    where $\langle ... \rangle_p$ denotes averaging over all points in the sample.
    
    \subsection{ Pairwise  distances} The  distribution  of  pairwise  distances  for  sets with  integral  dimensionality  is discussed  in  Ref.  20.
    
    \begin{equation}
	    f(l)=Dl^{D-1}(L/2)^{-D}I_{\mu}( \frac{D+1}{2},\frac{1}{2}),
    \end{equation}

    where $D$ is the integral dimensionality of the set, $l$ is the distance between a pair of points, $L$ is the greatest distance within  the  set, and $I_\mu(p,q)$ is the  incomplete  Bessel function  with $\mu=1-l^2/L^2$.  For $l<<L$ there  is  an  asymptote
    
    \begin{equation}
	    f(l)\sim l^{D-1},
    \end{equation}

    which  also  is  retained  for  sets  with  a  fractional  dimensionality  [21,22,14].

    \begin{figure*}[h!] \centering
	    \includegraphics[scale=0.5,clip]{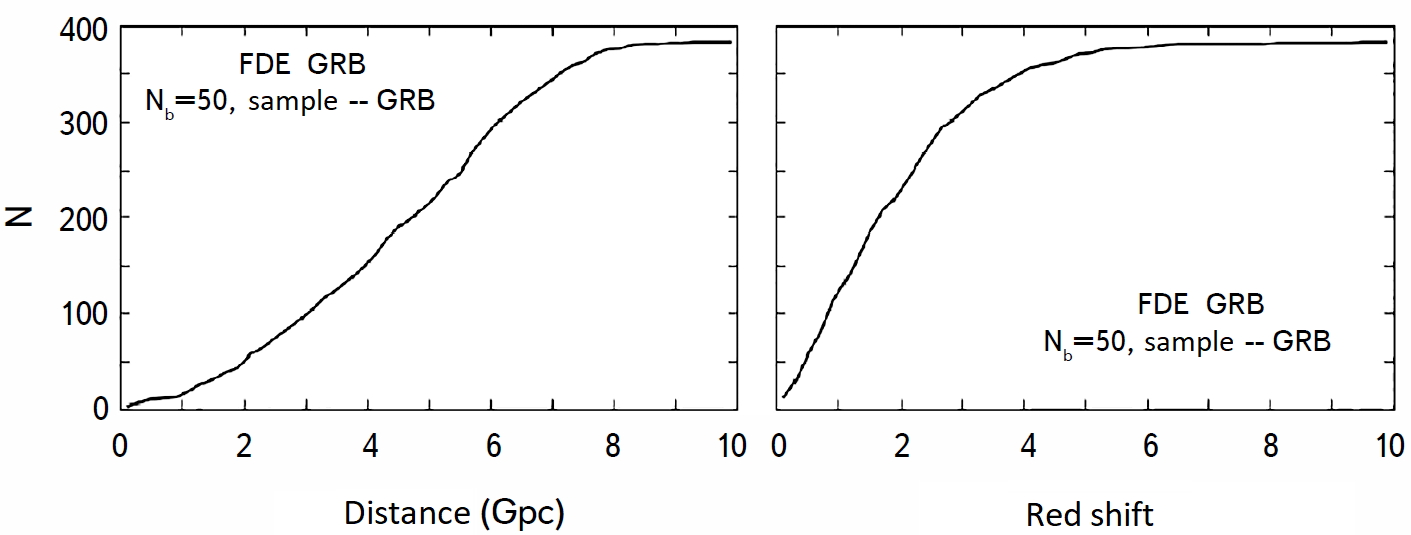}
	    \caption{Integrated distributions of the GRB with respect to distance ($\Delta R=200 Mpc$) and red shift ($\Delta z=0.2$).}
	    \label{fig:ID}
    \end{figure*}
    
        \begin{figure*}[h!] \centering
	    \includegraphics[scale=0.5,clip]{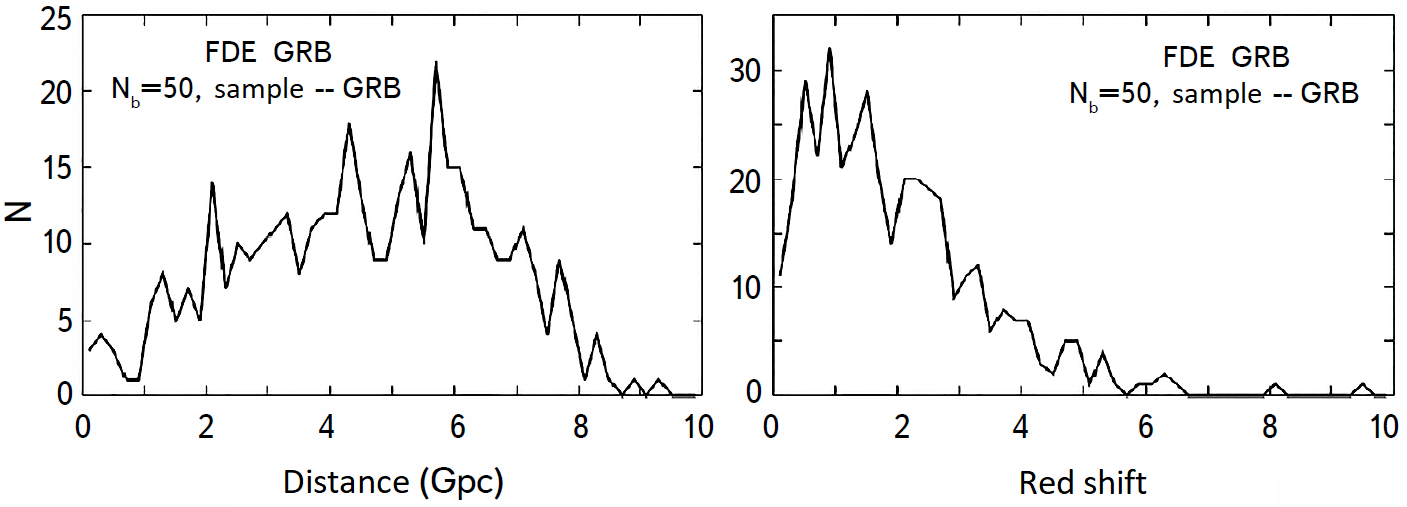}
	    \caption{Differential distributions of the GRB with respect to distance ($\Delta R=200 Mpc$) and red shift ($\Delta z=0.2$).}
	    \label{fig:DD}
    \end{figure*}
    
        \begin{figure}[h!] \centering
	    \includegraphics[scale=0.32,clip]{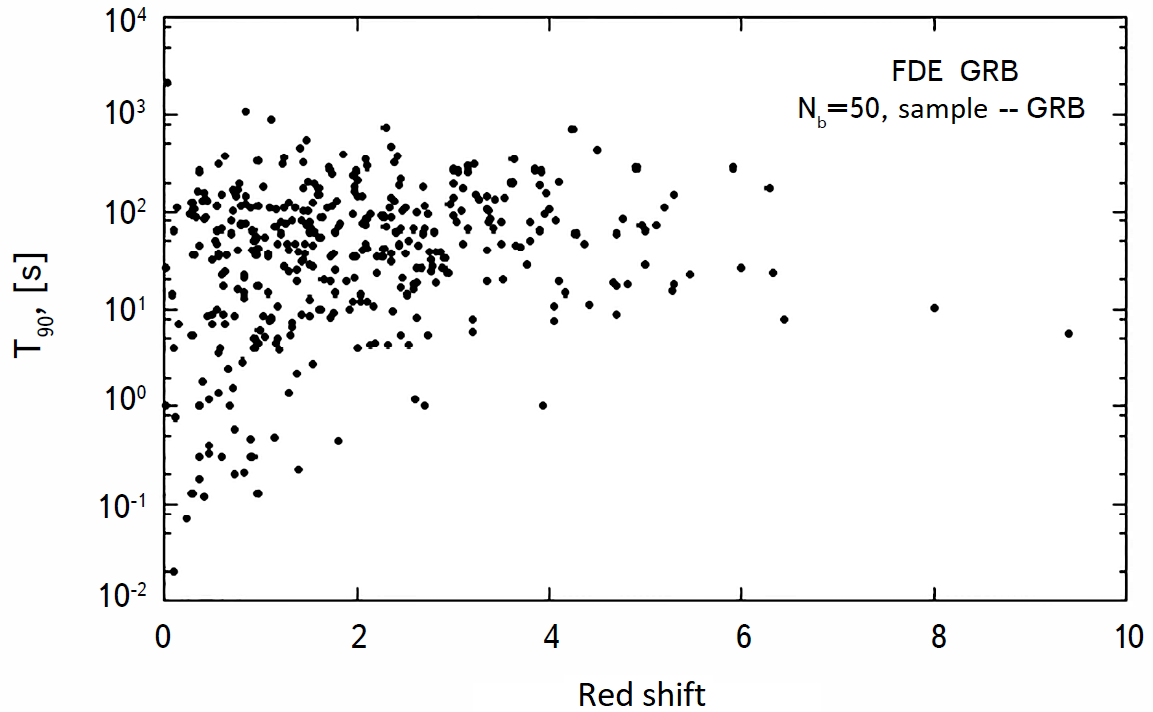}
	    \caption{The distribution of $T_{90}$ with respect to red shift $z$ for all the GRB sources}
	    \label{fig:T90}
    \end{figure}
    
        \begin{figure}[h!] \centering
	    \includegraphics[scale=0.36,clip]{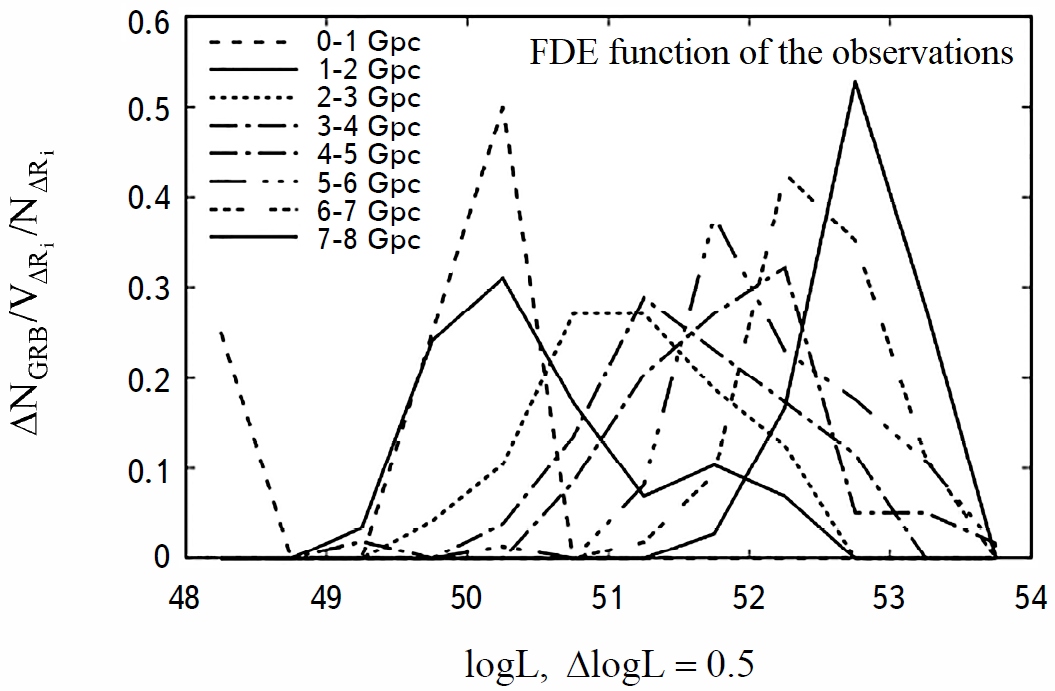}
	    \caption{ Observed  luminosity  distributions  of  the  GRB sources in fixed distance intervals.}
	    \label{fig:Lum}
    \end{figure}
    
    \subsection{Comparison of the methods} For a convenient comparison of the results of the pairwise distances and conditional density methods, plots of the ratio of the curves for a fractal or GRB to a uniform sample are ultimately examined.    In  this  configuration  the  slope  of  the  working  segment,  which  is  determined  by  the  condition  of  least squares  for  the  deviations,  is  equal  to  $D-3$ for  each  of the  methods  in  logarithmic  coordinates.
    
    For sparse sets with $N<10^3$ the distributions of the conditional density and pairwise distances are characterized by  substantial  random  deviations  from  a  power  law  dependence.    To  reduce  the  influence  of  this  effect,  the  curves are averaged over a sufficient number of runs for the given set.  This is done by specifying different zero points for the  pseudorandom  number  generator.  This  operation  is  carried  out  for  the  conditional  density  distributions,  as  well as  for  the  pairwise  distance  distributions.
    
    \subsection{Calculating the distance and luminosity} In the standard cosmological model the metric distances are given  in  terms  of  red  shift  by  the  formula  [4]
    
    \begin{equation}
	    R(z)_{Mpc}=\frac{c}{H_0} \int_0^z (\Omega^0_v+\Omega_m^0(1+z')^3-\Omega_k^0(1+z')^2)^{-1/2} dz',
    \end{equation}

    where $H_0=70$ km/s/Mpc  is the Hubble  constant, $c=3\cdot10^{10}$ cm/s is  the  speed of light, $\Omega^0_v=0.7$, $\Omega_m^0=0.3$, and $\Omega_k^0=0$ are  the  cosmological parameters,  and  $z$  is  the  red  shift.    Then  the  spherical  coordinates  are  transformed  to Cartesian,  since  the  space  is  Euclidean.  The  luminosity  in  terms  of  the  $\Lambda CDM$ model  is  given  by
    
    \begin{equation}
    	L(z)=4\pi S_{obs} R(z)_{sm}^2 (1+z)^2, 
    \end{equation}

    where $S_{obs}$ erg/s/cm$^2$ is the GRB flux, which equals the ratio of the radiation received in the $15\div150$ keV range to the  time $T_{90}$,  and  $z$  is  the  red  shift.
    
    \subsection{The  model  catalogs} Fractal  and  uniform  sets  analogous  to  the  GRB catalog  were  modeled  for comparison  with  the  actual  sample.    Since  strong  absorption  of  visible  radiation  (leading  to  additional  selection  of GRB when measuring their red shift) occurs near the galactic belt, it is necessary examine the influence of this effect on  the  estimated  fractal  dimensionality.    Thus,  two  cases  were  examined.    Examples  of  these  are  shown  in  Fig.  6. The first covers the entire celestial  sphere,  while  the band from $-10$ to $+10$  degrees of galactic latitude  is cut out in the  second.    All  of  these  samples  are  bounded by  a  sphere  of  radius  8  Gpc.
    
    \begin{figure}[h!] \centering
	    \includegraphics[scale=0.36,clip]{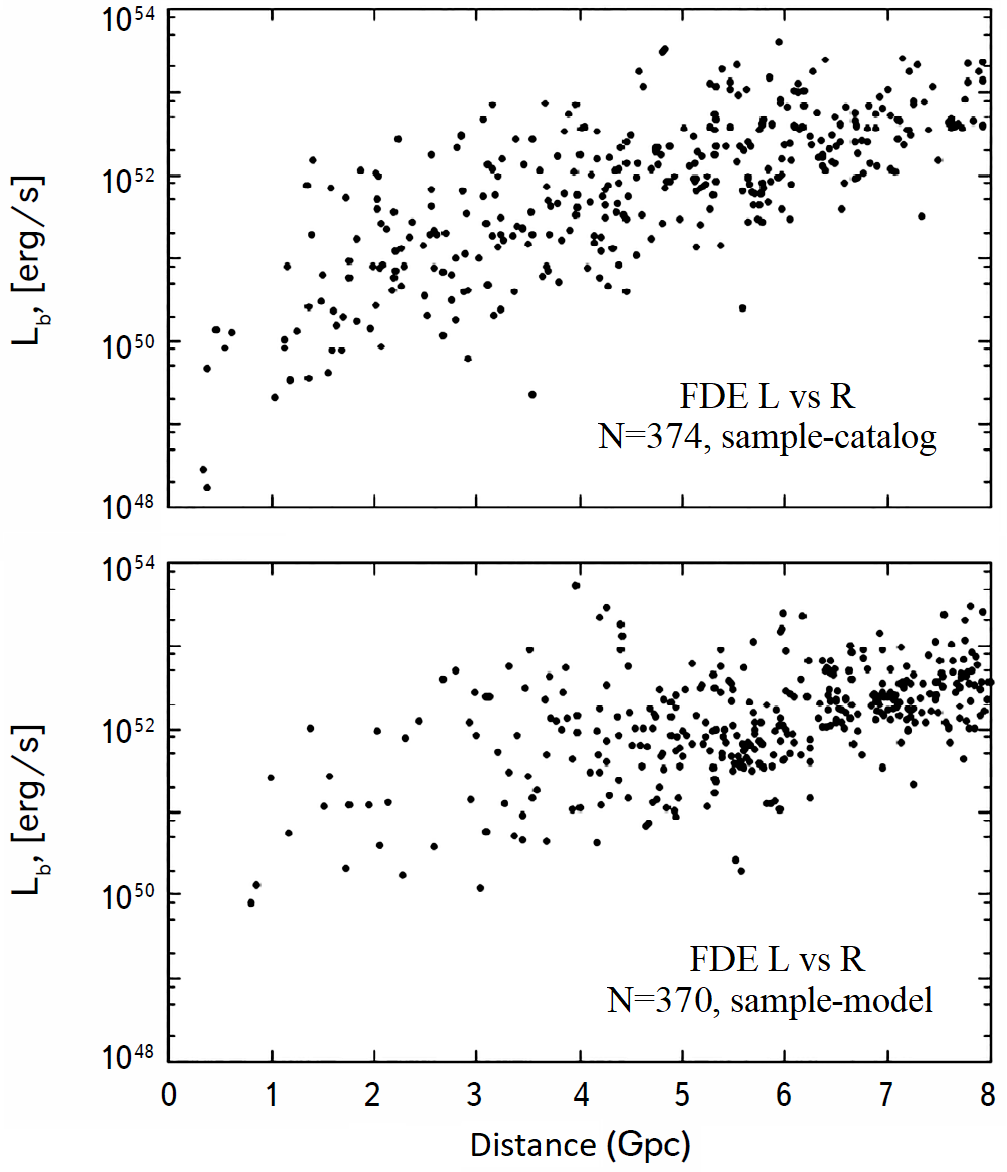}
	    \caption{Distribution of the luminosities of the GRB catalog and the uniform model catalog with respect to distance.}
	    \label{fig:LvsR}
    \end{figure}
    
    To determine the fractal dimensionality  over the entire volume right away, it  is  necessary  to account for the observed  selection  effects,  e.g.,  the  Malmquist  effect,  for  all  the  points.    For  this,  the  observed  profiles  in  spherical layers with a step size of 1 Gpc shown in Fig. 4 are taken as the model visible luminosity function.  Since introducing a model  selectivity  with respect  to  luminosity affects  the  conditional  density  distribution,  the  ratio  of  the  curves for fractals  and  GRB  to  the  uniform  curves  is  examined.

    \section{Results} 
    
    \subsection{General properties of the GRB source catalog} The radial distributions of the GRB catalog are shown in  Fig.  1  (integrated  distributions)  and  Fig.  2  (differential  distributions).  The  distribution  of  the  time  an  event  is observed, $T_{90}$, with respect to red shift is shown in Fig. 3.  At present, the drift with increasing red shift toward shorter event  times  predicted  by  the  standard  model  has  not  yet  been  observed.    Our  result  is  consistent  with  Ref.  23.  A definitive answer to this question would require a significant increase in the  number  of  GRB at large red shifts and it  would  be  necessary  to  study  the  dependence  of  $T_{90}$  on  the  distance  to  the  GRB  source.

    GRB  can  serve  as  an  indicator  of  galactic  clusters,  so  it  is  possible  look for  close  pairs  in  the  spatial distribution of GRB in Table 3.  Of the 18 pairs, it is possible to identify spatially distinctive structures of three and four  gamma-ray  bursts,  as  well  as  two  pairs  for  which  the  distance  between  the  sources  is  less  than  100  Mpc  at $z\approx0.013$ and $z\approx1.43$.
    
    \begin{table}[h!] \centering \footnotesize
    	\caption{ GRB  Sources  with  Pairwise  Distances  to  300  Mpc.}
		\begin{tabularx}{.48\textwidth}{>{\hsize=0.08\textwidth}ccrrrc}
			\hline	\hline 			
			N	&	Designation	&	$d_{Mpc}$	&	\centering$l$	&	$b$	&	$z$	\\
			\hline 
	175	&	111005A	&	79.4	&	338.33759	&	34.63886	&	0.013	\\
	4	&	100316D	&			&	266.91664	&	--19.78007	&	0.014	\\
	
	175	&	111005A	&	195.6	&	338.33759	&	34.63886	&	0.013	\\	
	158	&	060218A	&			&	166.86303	&	--32.86884	&	0.033	\\
		
	4	&	100316D	&	150.1	&	266.91664	&	--19.78007	&	0.014	\\
	158	&	060218A	&			&	166.86303	&	--32.86884	&	0.033	\\
		
	158	&	060218A	&	293.1	&	166.86303	&	--32.86884	&	0.033	\\
	94	&	051109B	&			&	100.54662	&	--19.39992	&	0.080	\\
		
	234	&	060505A	&	264.9	&	22.09128	&	--53.71345	&	0.089	\\
	54	&	060614A	&			&	344.08607	&	--43.94594	&	0.130	\\
	
	32	&	061201A	&	206.0	&	315.71506	&	--38.23391	&	0.111	\\
	54	&	060614A	&			&	344.08607	&	--43.94594	&	0.130	\\
		
	117	&	130427A	&	261.1	&	206.48629	&	72.51440	&	0.340	\\
	18	&	130603B	&			&	236.47527	&	68.43758	&	0.356	\\
			
	13	&	110328A	&	299.3	&	86.71625	&	39.42626	&	0.354	\\
	84	&	151027A	&			&	90.49260	&	28.48382	&	0.380	\\
		
	25	&	140903A	&	276.7	&	44.40465	&	50.11996	&	0.351	\\
	89	&	101213A	&			&	37.17510	&	45.89511	&	0.414	\\
			
	27	&	070724A	&	282.8	&	184.32601	&	--73.81347	&	0.457	\\
	315	&	091127A	&			&	197.38677	&	--66.73665	&	0.490	\\
			 
	22	&	141212A	&	242.7	&	155.24497	&	--38.01808	&	0.596	\\
	304	&	130215A	&			&	163.06996	&	--39.75075	&	0.597	\\
			 
	104	&	150323A	&	189.0	&	174.83011	&	36.29286	&	0.593	\\
	168	&	110106B	&			&	172.91003	&	40.47816	&	0.618	\\
			
	291	&	080916A	&	236.0	&	333.57758	&	--50.49415	&	0.689	\\
	124	&	150821A	&			&	329.53474	&	--52.37413	&	0.755	\\
			
	161	&	050824A	&	248.8	&	122.21283	&	--40.26635	&	0.830	\\
	77	&	080710A	&			&	116.98198	&	--43.17503	&	0.845	\\
			
	77	&	080710A	&	291.4	&	116.98198	&	--43.17503	&	0.845	\\
	271	&	060912A	&			&	113.49717	&	--41.34504	&	0.937	\\
			 
	206	&	160131A	&	231.1	&	207.86277	&	--25.13766	&	0.970	\\
	116	&	120907A	&			&	208.46983	&	--29.19799	&	0.970	\\
			
	50	&	161108A	&	269.0	&	221.80332	&	78.92167	&	1.159	\\
	268	&	90530	&			&	212.46605	&	77.98286	&	1.266	\\
			
	48	&	050822X	&	77.0	&	255.27955	&	--54.45517	&	1.434	\\
	203	&	050318A	&			&	256.44382	&	--55.23286	&	1.440	\\
			\hline 	
		\end{tabularx}
		\label{tab:paires}				
\end{table}
    
    \subsection{Fractal  dimensionality} Figure  5  is  a  comparison  of  the  luminosity distribution  of  the  GRB  with  the model uniform catalog after application of the selectivity function with respect to luminosity shown in Fig. 4.  Figure6 illustrates  the  visible  differences between the real sample of GRB and the model catalogs for two geometries.    In the case of the truncated celestial sphere, the sample is a superposition of two hemispheres.  Plots of the conditional density  and  mutual  distances  are  shown  in  Figs.  7  and  8,  respectively.    The  graphs  of  the  fractals  shown  here  are average curves over 17  runs  and this is the  cause  of the corresponding deviations for the points  in  the graphs.  The amplitude of the deviations in the pairwise distance method is a factor of two larger than for the conditional density.In  all  runs  of  the  fractal  and  uniform  samples  to  which  these  methods  were  applied  directly,  the  number  of  points is  roughly  equal  to  the  number  of  GRB.    This  is  done  by  creating  an  excess  of  points  in  the  initial  run,  and  then by  selecting  them  uniformly.
    
    \begin{figure*}[h!] \centering
	    \includegraphics[scale=0.5,clip]{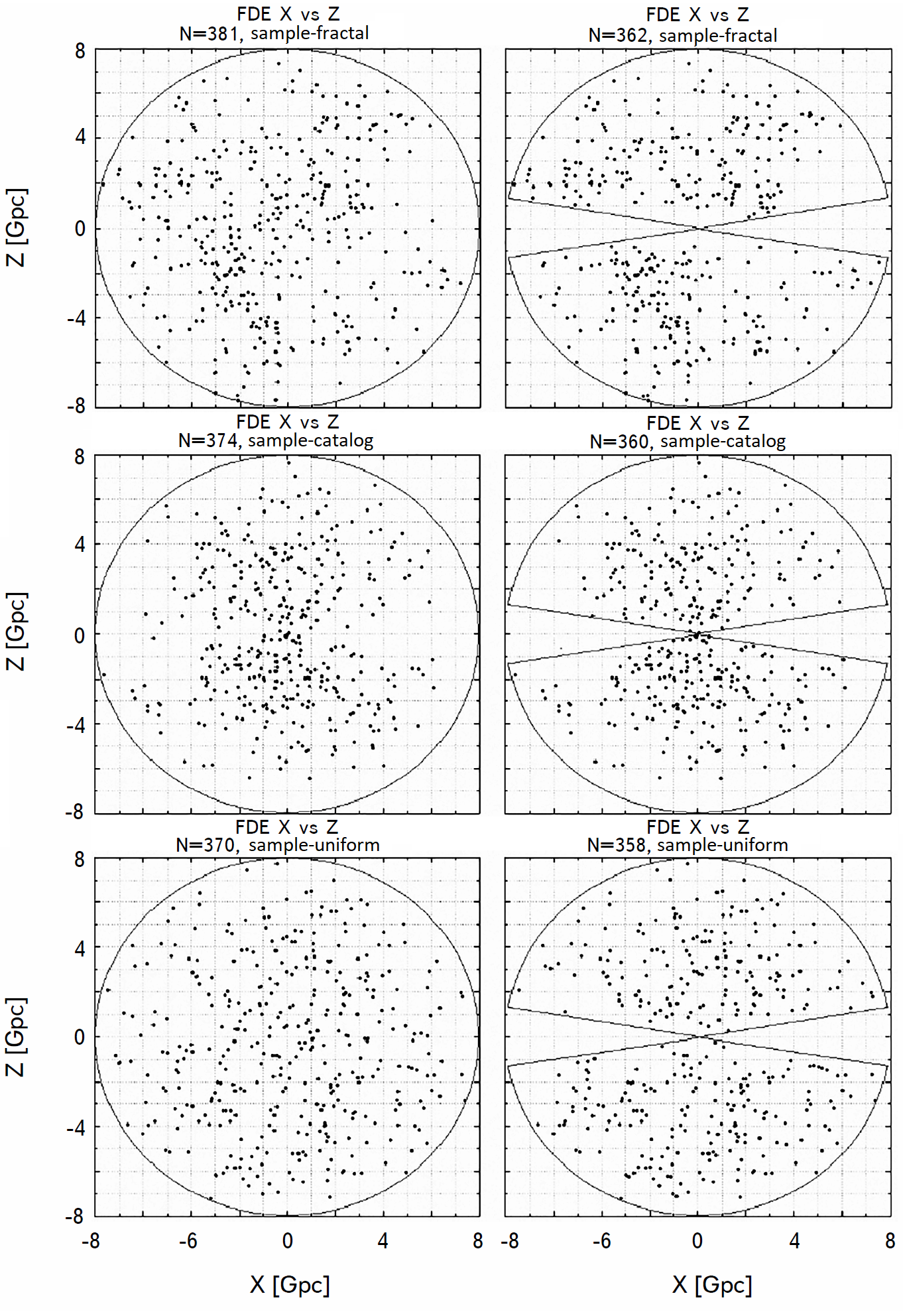}
	    \caption{Projections of the spatial distribution on the X-Z plane for two geometrical shapes of the sample.  The top corresponds to a fractal $D=2.5$, the center to the GRB catalog, and the bottom to a uniform set.}
	    \label{fig:Projs}
    \end{figure*}
    
    For a more accurate result, the excess in the observed number of GRB on  small radial scales must be taken into  account,  as  discussed  in  Ref.  24.
    
    \begin{figure}[h!] \centering
	    \includegraphics[scale=0.28,clip]{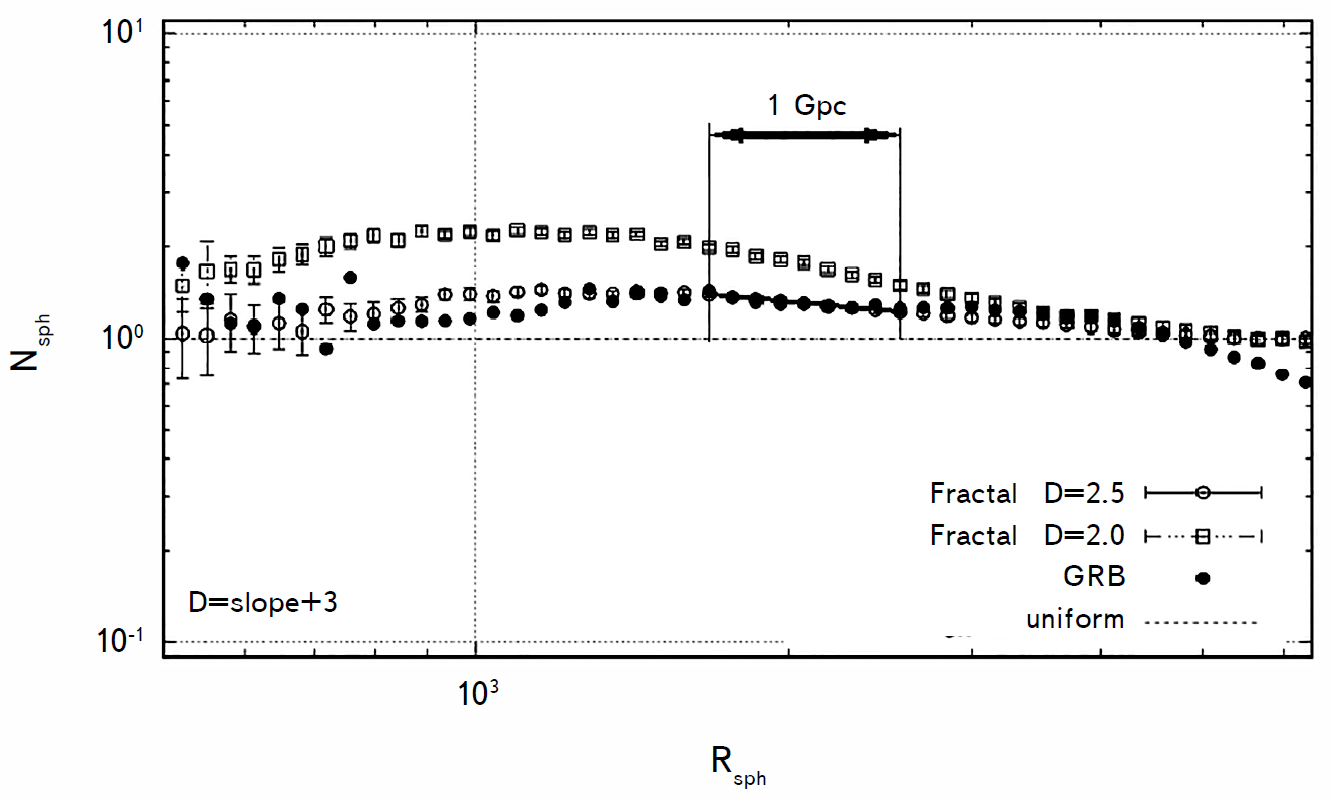}
	    \caption{Plots of the normalized conditional density for GRB (solid circles) and fractals $D=2.0$ (squares) and $D=2.5$ (circles) for the full celestial sphere.  Unity corresponds to a uniform distribution.}
	    \label{fig:CD}
    \end{figure}
    
    \begin{figure}[h!] \centering
	    \includegraphics[scale=0.28,clip]{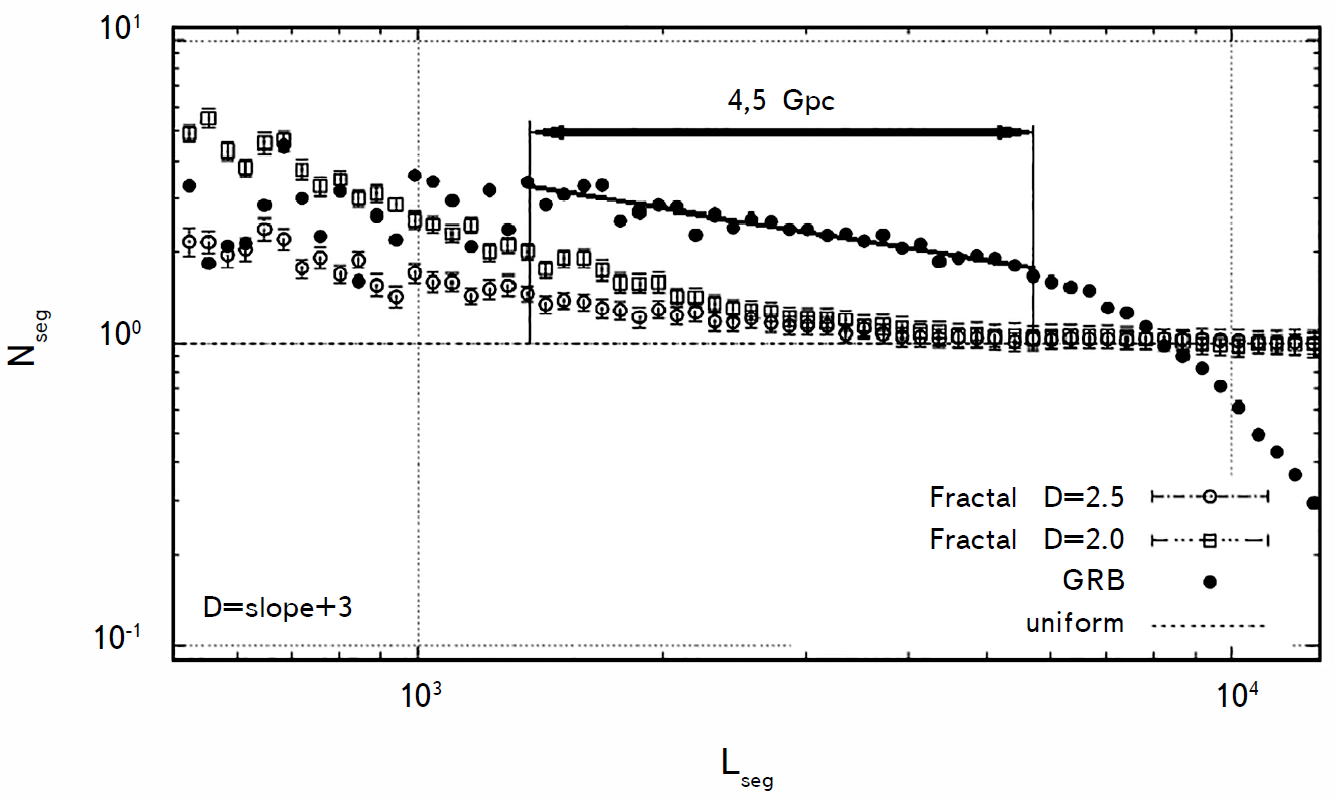}
	    \caption{ Plots  of  the  reduced  pairwise  distances  for  GRB  (solid  circles) and fractals $D=2.0$ (squares) and $D=2.5$ (circles) for the truncated celestial sphere.Unity corresponds to a uniform distribution.}
	    \label{fig:PD}
    \end{figure}    
    
    \section{Conclusions} For model sets with $N>10^3$ points, the accuracy of the estimated fractal dimensionality reaches $\Delta D=0.06$ for  the  conditional  density and $\Delta D=0.03$ for  pairwise  distances  prior  to  introduction  of  the  model luminosity function  and  prior  to  truncation  of  the  galactic  belt.    After  introducing  limitations  on  the  geometry  and  luminosity with the same total number of points, the accuracy of the method decreased by a factor of two.  It is noteworthy that the  radius  of  the  sphere,  and  not  the  diameter,  is  treated  as  the  parameter  in  the  conditional  density,  so  that  the horizontal axis can be multiplied by a factor of two when comparing the methods.  The main parameters determining the accuracy of the estimated fractal dimensionality are the number of points and the number of runs.  During testing of the accuracy, it was found that the pairwise distance method is better than the conditional density method for short scales.
    
    Thus, the pairwise distance curve reaches a power law dependence at a scale equal to twice the size of the elementary cell of the fractal structure. At the same time the conditional density begins to operate at a scale that is $4\div8$ times the size of the elementary cell, depending on the representativeness of the sample and the value of the fractal dimensionality. On large scales both methods have problems. In the conditional density, averaging is over a small number of spheres, while in the pairwise distances the distribution behaves unpredictably or is close to uniform.
    
    For the case of a full celestial sphere, the estimated fractal dimensionality of the distribution of the GRB sources was $D=2.6\pm0.12$ at $R=1.5\div2.5$ ($d=3\div5$) Gpc for the conditional density and $D=2.6\pm0.06$ for $l=1.5\div5.5$ Gpc. In the case with a truncated galactic belt, the conditional density does not yield a unique result and its distribution does not differ from uniform. The pairwise distances have a stable power law dependence with $D=2.6\pm0.06$ and essentially do not change the interval for the linear segment $l=1.5\div5.5$. Thus, on scales of $\approx 3\div5$ Gpc, there is a power law correlation between the methods.
    
    At distances close to $D=3$, when selection effects are taken into account the methods yield a systematically high estimate for the model sets by roughly $\Delta D=0.1$. Given this bias, it is necessary to correct the result in order to obtain a more accurate estimate. Thus, the estimated fractal dimensionality of the observed GRB distribution is $D=2.55\pm0.06$.
    
%====================================================

\end{document}